\documentstyle[graphics]{aa}

\begin{document}


\title{XMM-Newton Detection of a Comptonized Accretion Disc in the Quasar 
PKS~0558$-$504}
\author{P.T. O'Brien\inst{1}
\and J.N. Reeves\inst{1}
\and M.J.L. Turner\inst{1}
\and K.A. Pounds\inst{1}
\and M. Page\inst{2}
\and M. Gliozzi\inst{3}
\and W. Brinkmann\inst{3}
\and J.B. Stephen\inst{4}
\and M. Dadina\inst{4}}

\offprints{P.T. O'Brien}
\mail{pto@star.le.ac.uk}

\institute{X-ray Astronomy Group, Department of Physics and Astronomy,
University of Leicester, LE1 7RH, U.K.
\and MSSL, University College London, Holmbury St. Mary, Dorking, Surrey,
RH5 6NT, U.K.
\and Max-Plank-Institut f\"{u}r extraterrestrische Physik, Postfach 1312,
D-85741, Garching, Germany
\and Istituto TESRE, CNR, Via Gobetti 101, I-40129 Bologna, Italy}

\date{Received September 2000 / Accepted October 2000}

\maketitle

\begin{abstract}
We present {\it XMM-Newton} observations of the bright quasar
PKS~0558$-$504. The 0.2--10 keV spectrum is dominated by a large,
variable soft X-ray excess. The fastest flux variations imply
accretion onto a Kerr black hole. The {\it XMM-Newton} data suggest
the presence of a `big blue bump' in PKS~0558$-$504 extending from the
optical band to $\sim 3$~keV. The soft X-ray spectrum shows no
evidence for significant absorption or emission-line features. The
most likely explanation for the hot big blue bump is Comptonization by
the multi-temperature corona of a thermal accretion disc running at
a high accretion rate.
\end{abstract}

\section{Introduction}

The most prominent feature in the broad-band spectral energy
distribution of many active galactic nuclei (AGN) is the `big blue
bump' (BBB), which contains a large fraction of the bolometric
luminosity. In most broad-line AGN the BBB begins around 1 $\mu$m,
rises to a peak in the ultraviolet and then declines again into the
unobservable EUV (e.g. Elvis et al. \cite{el}). The soft X-ray excesses
seen in many AGN may be the high-energy tail of the BBB (e.g. Turner
\& Pounds \cite{tutwo}; Walter et al. \cite{wa}; Laor et al. \cite{la}). The harder
2--10~keV X-ray spectrum can usually be quite well described by a
power-law with photon-index $\Gamma \approx 1.9$. Most AGN also
show an iron K$\alpha$ fluorescence feature around 6--7 keV and/or a
hump of emission above 10 keV (e.g. Pounds et al. \cite{pou}), widely
interpreted in terms of Compton reflection of hard X-rays by
optically-thick matter probably in the form of an accretion disc
(George
\& Fabian \cite{ge}).

Three basic classes of model have been proposed to explain the BBB.
The first class involves thermal emission from an optically thick
accretion disc surrounding the central black hole in which the primary
heating is by viscous dissipation within the disc itself (e.g. Shields
\cite{sh}). The disc thermal spectrum may be comptonized by a hot disc
corona; a process which changes the observed shape of the continuum,
helping to explain soft X-ray emission from discs which are otherwise
too cool (e.g. Czerny \& Elvis \cite{cz}). A second class of models also
involves thermal emission from an accretion disc, but with irradiation
by a hard X-ray power-law (perhaps arising from magnetic flares) as
the primary energy source (e.g. Ross \& Fabian \cite{ro}). In these models
the disc reprocesses the hard X-ray emission into thermal emission
which is re-radiated at the local disc temperature. Irradiated discs
may have flatter radial temperature profiles, and hence different
spectra, compared to discs dominated by viscous dissipation. The final
class of model involves optically thin, thermal free-free emission
from a hot plasma (Barvainis \cite{ba}), although single temperature models
of this kind do not fit the observed continuum very well
(Siemiginowska et al. \cite{si}).

While the spectral form of the optical and UV continuum in AGN are
well constrained by existing data, the soft X-ray continuum shape is
often quite poorly defined for individual QSOs and therefore places
only limited constraints on BBB models. These limitations have been
due to the low spectral resolution, signal-to-noise ratio and/or
narrow spectral bandpass provided by previous X-ray instruments,
thereby making it difficult to quantify absorption/emission features
which may contaminate the X-ray spectrum and hence distort the derived
continuum shape. The advent of {\it XMM-Newton} with its broad
bandpass, high throughput and good spectral resolution now provides
the opportunity to obtain high-quality X-ray spectra and hence make
progress in defining the intrinsic shape of the X-ray continuum in
AGN.

In this paper we present {\it XMM-Newton} observations of 
PKS~0558$-$504, a low-redshift quasar ({\it z} $= 0.137$;
$M_{\mathrm{V}} \approx -24.5$) identified during the {\it HEAO-1} survey. It
has quite narrow Balmer lines (FWHM $= 1500$ km s$^{-1}$, Remillard et
al. \cite{re}), which has led to it being routinely described as a Narrow-Line
Seyfert 1 (NLS1), although we note that the UV lines, such as Ly$\alpha$,
are broader (FWHM $\approx 3800$ km s$^{-1}$; Kuraszkiewicz et al.
\cite{ku}).
PKS~0558$-$504 is unusual in being one of the few radio-loud NLS1-type
galaxies (Schartel et al. \cite{sch}). We find that the X-ray spectrum of
PKS~0558$-$504 is dominated by a strong soft X-ray excess, which may be
the high-energy tail of the BBB. The X-ray spectrum is not contaminated
by either cold or warm (ionised) absorption features. Soft X-ray narrow
emission-lines are also excluded at a high significance level. Thus, we
claim to have detected the featureless X-ray continuum emission from a
luminous quasar.

\begin{figure}
\resizebox{\hsize}{!}{\includegraphics{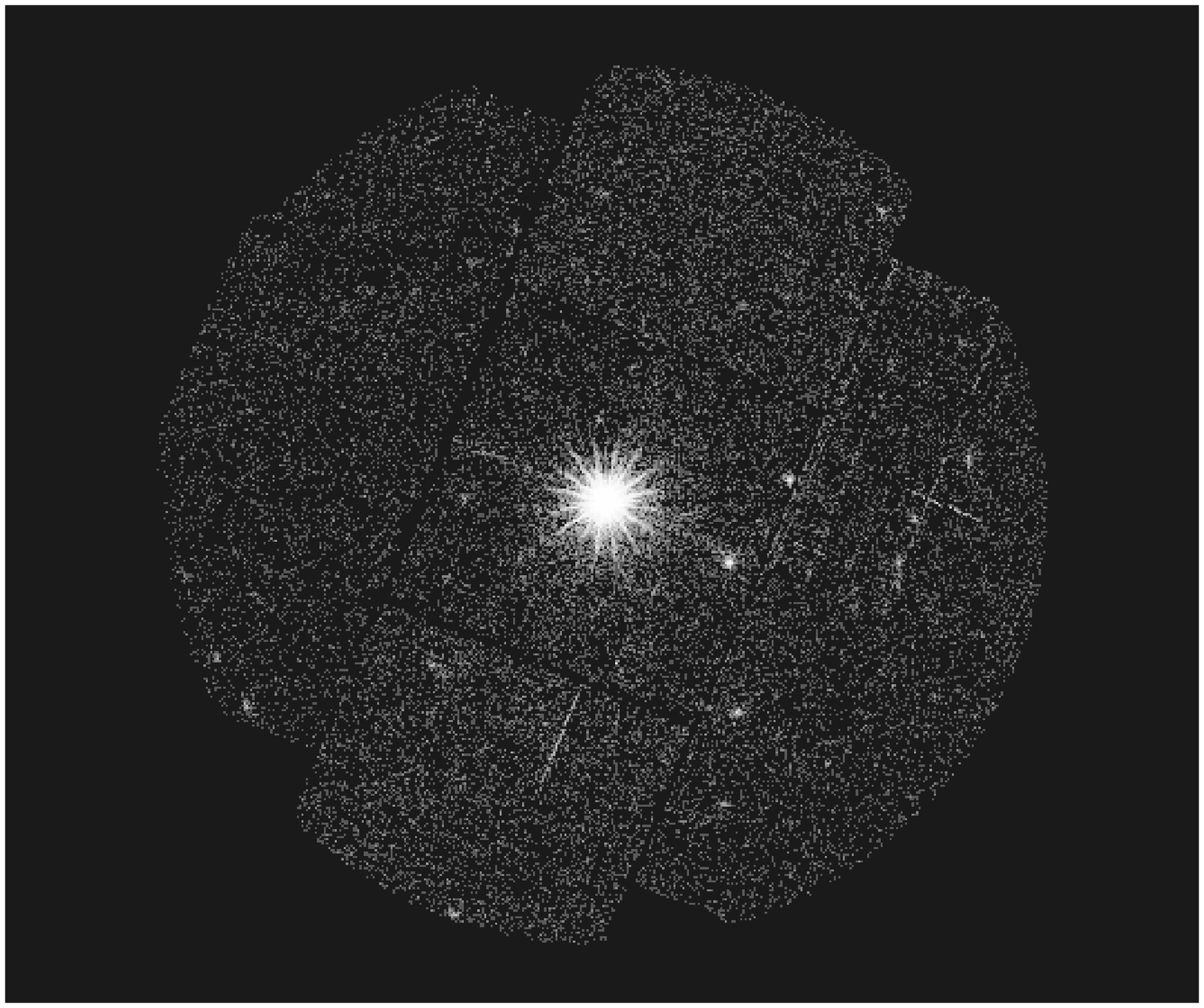}}
\caption{A Full-Window mode EPIC MOS-1 image of PKS~0558$-$504 taken
during {\it XMM-Newton} orbit 42. A logarithmic intensity scale has been
used to show the point spread function. No other strong X-ray sources are
seen within the 30 arcminute field of view.}
\label{mosimage}
\end{figure}

\section[]{XMM-Newton Observations}

We present observations of PKS~0558$-$504 taken with the European
Photon Imaging Camera (EPIC) during orbits 42 and 84 (1--2 March 2000
and 24--25 May 2000) of the calibration and performance verification
phase of {\it XMM-Newton}. The EPIC instrument is described in Turner
et al. (\cite{tuone}) and Strueder et al. (\cite{str}). During both
orbits the EPIC MOS cameras were operated in several different
observing modes. Here we include only those observations taken with
the medium filter, for which the total EPIC MOS exposure times for
orbits 42 and 84 are 85.5 ksecs and 53.7 ksecs respectively. Only the
full-frame mode EPIC PN data for orbit 84 were used, for which the
total exposure time is 10.2 ksecs.

For operational reasons, the EPIC observations in each orbit were
divided into several separate exposures of varying lengths. These
exposures were reduced using the {\it XMM-Newton} {\sc SAS} (Science
Analysis Software) package. The event lists output from the
standard {\sc emchain} and {\sc epchain} scripts were further filtered
for each EPIC exposure using the {\sc SAS} {\sc xmmselect} task. For the
MOS exposures only X-ray events corresponding to patterns 0--12 (similar
to {\it ASCA} event grades 0--4) were selected. Only pattern 0 (single
pixel) events were selected for the PN. Known hot or flickering pixels
and electronic noise were rejected using the {\sc SAS} and the low-energy
cutoff was set to 200 eV. One of the EPIC MOS-1 images from orbit 42
is shown in Fig.\ \ref{mosimage}.

Source spectra and light curves were extracted from each EPIC exposure
using a circular region of diameter 1 arcminute centred on the source
position. PKS~0558$-$504 is by far the brightest X-ray source in the
30 arcminute EPIC field-of-view. The EPIC cameras are subject to non
X-ray background events due to particles which can vary significantly
during an exposure. Background spectra and light curves for use in
later data analysis were derived in a similar way to the source data
using an offset region close to the source. We note that the observed
count-rate for PKS~0558$-$504 implies little effect due to pile-up and
no significant spectral differences were found between observations
taken in different MOS window modes. The {\sc Xspec v11.0} software
package was used to calibrate the background-subtracted EPIC spectra
using the most recent camera response matrices derived from
ground-based and in-orbit data. Before further analysis the EPIC
spectra were binned to give a minimum of 20 counts per bin.

Reflection Grating Spectrograph (RGS; den Herder et al. \cite{den})
exposures from orbit 84 were also reduced to search for weak/narrow
soft X-ray absorption or emission-line features. Three exposures were
analysed, two from RGS-1 (each of 25 ksecs) and one from RGS-2 (31
ksecs). The RGS data were first processed using the {\sc SAS} {\sc
rgsproc } script. The merged RGS event lists were then fed to the {\sc
SAS} {\sc evselect} and {\sc xmmselect} tasks, and both source and
background spectra for each exposure were extracted interactively.
These spectra were then calibrated using response matrices derived
from the individual RGS exposures using the {\sc SAS} {\sc rmfgen}
task.

The Optical Monitor (OM; Mason et al. \cite{ma}) imaging data for orbits 42
and 84 were reduced using standard {\sc SAS} tasks and calibrated with
the instrument response information for each filter.

\begin{figure}
\resizebox{\hsize}{!}{\rotatebox{-90}{\includegraphics{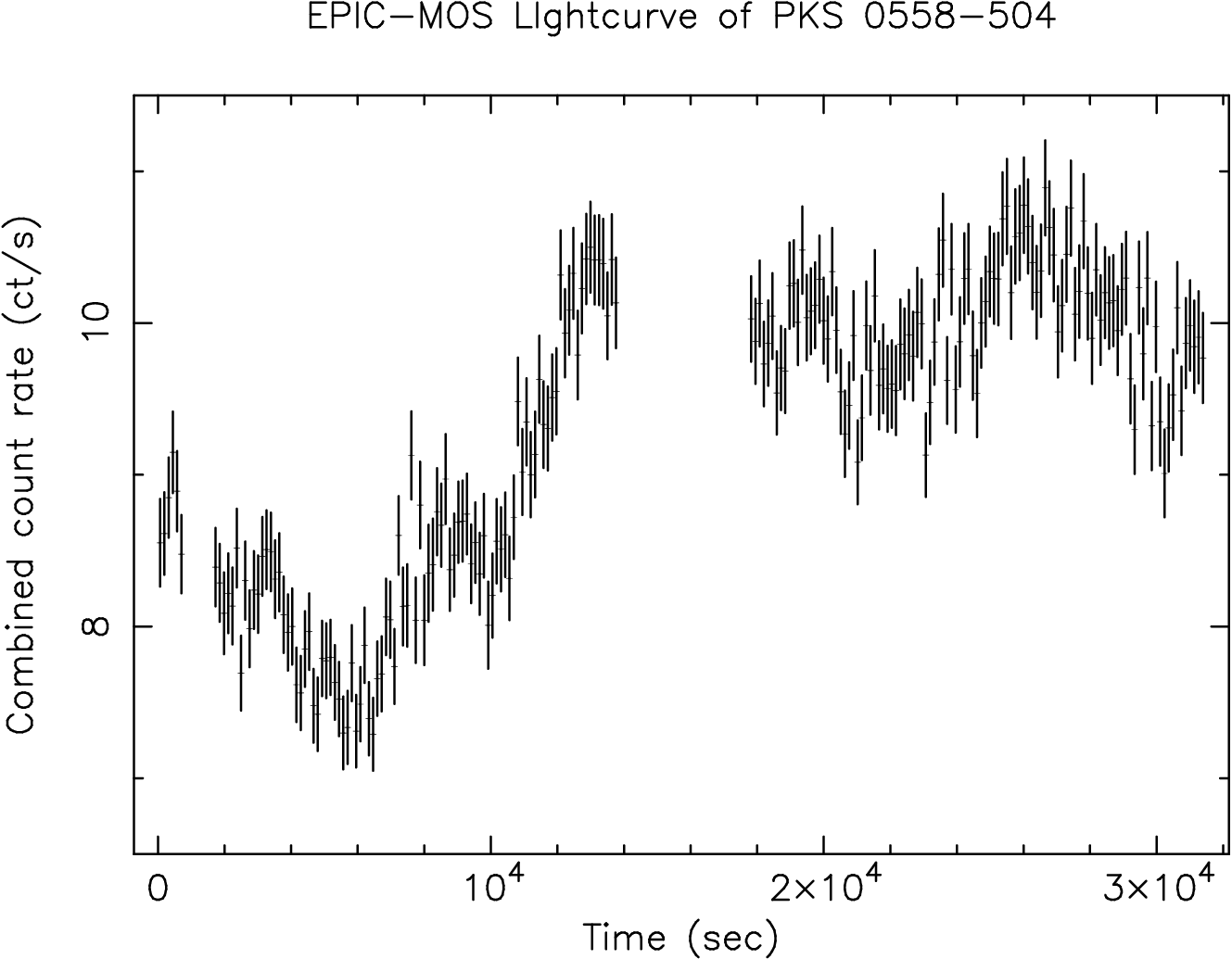}}}
\caption{The EPIC MOS-1 + MOS-2 0.2--10 keV
light curve of PKS~0558$-$504 for orbit 84 in
125 second time bins.}
\label{lcurve}
\end{figure}

\section{Variability}

No substantial variability was detected within individual EPIC exposures
on timescales less than a few minutes, so the data were rebinned and
assembled into orbit light curves by combining the individual
light curves for each exposure in order to search for longer duration
events. The PKS~0558$-$504 broad-band EPIC MOS light curve for orbit 84,
binned into 125 seconds bins, is shown in Fig.\ \ref{lcurve}.

Strong broad-band X-ray flux variability is seen during both orbits, with
similar rates of increase and decrease. The largest change in count rate
occurs during orbit 84 -- an increase of 35\% in 3000 seconds.
Based on the spectral fits discussed below and using $H_0 = 50$ km
s$^{-1}$ Mpc$^{-1}$ and $q_0 = 0.5$, this corresponds to a rate of
increase in the 0.2--10 keV luminosity of $\Delta {\mathrm{L}}/\Delta
{\mathrm{t}} = 5 \times 10^{41}$ erg s$^{-2}$.
Assuming photon diffusion through a spherical mass of accreting matter
in which the opacity is dominated by Thomson scattering, the observed
luminosity change implies a minimum efficiency of converting matter
into energy of $\eta = (\Delta {\mathrm{L}}/\Delta {\mathrm{t}}) / (2
\times 10^{42})$ (Guilbert, Fabian \& Rees \cite{gu}). In the
simplest interpretation, the change in luminosity observed during
orbit 84 therefore implies $\eta = 0.25$, an efficiency which exceeds
the theoretical maximum for a non-rotating (Schwarzschild) black hole
but is consistent with that for a maximally rotating Kerr black hole.
Some contribution of beamed X-rays cannot be ruled out, but the shape
of the continuum (discussed below) is not consistent with that of a
typical beamed source. A tendency for the X-ray continuum to harden
slightly when brighter is seen when comparing the data from orbits 42
and 84. The X-ray variability of PKS~0558$-$504 is discussed in more
detail in a separate paper (Gliozzi et al. \cite{gltwo}).

During a {\it Ginga} observation, the PKS~0558$-$504 continuum
appeared to vary by 67\% in 3 minutes (implying $\eta \sim 2$;
Remillard et al. \cite{rem}). Given the high count-rates and well determined
background of the {\it XMM-Newton} observations, we could have
detected such a flare on timescales as short as a few tens of seconds,
but none were observed. {\it ROSAT} HRI observations also failed to
detect such rapid flaring (Gliozzi et al. \cite{gl}), but did show
luminosity variability rates (allowing for their use of $H_0 = 70$ km
s$^{-1}$ Mpc$^{-1}$) which imply accretion efficiencies similar to
that derived from the {\it XMM-Newton} data. Without confirmation
using an imaging detector it is impossible to confirm or deny the
association of the {\it Ginga} flare with intrinsic variability in
PKS~0558$-$504, so we do not consider the {\it Ginga\/} flare any
further in this paper.

\section[]{Spectral Analysis}

Previous analyses of {\it ASCA} observations of PKS~0558$-$504 imply
an X-ray spectrum which is softer at lower energies. Fixing the
absorbing column density at the Galactic value of $4.4 \times 10^{20}$
cm$^{-2}$, Gliozzi et al. (\cite{gl}) find $\Gamma = 2.99 \pm 0.09$ over
0.1--2.4 keV from {\it ROSAT} data. {\it ASCA} data give $\Gamma =
2.26 \pm 0.03$ over 0.6--10 keV (Vaughan et al. \cite{va}), whereas the
{\it Ginga} data reveal a typical AGN power-law spectrum with $\Gamma =
1.89^{+0.04}_{-0.09}$ over 2--18 keV (Lawson \& Turner \cite{law}).
Similarly, Comastri (\cite{com}) finds $\Gamma = 1.97\pm0.04$ over 2--10 keV
from {\it BeppoSAX\,} data.

\begin{figure}
\resizebox{\hsize}{!}{\rotatebox{-90}{\includegraphics{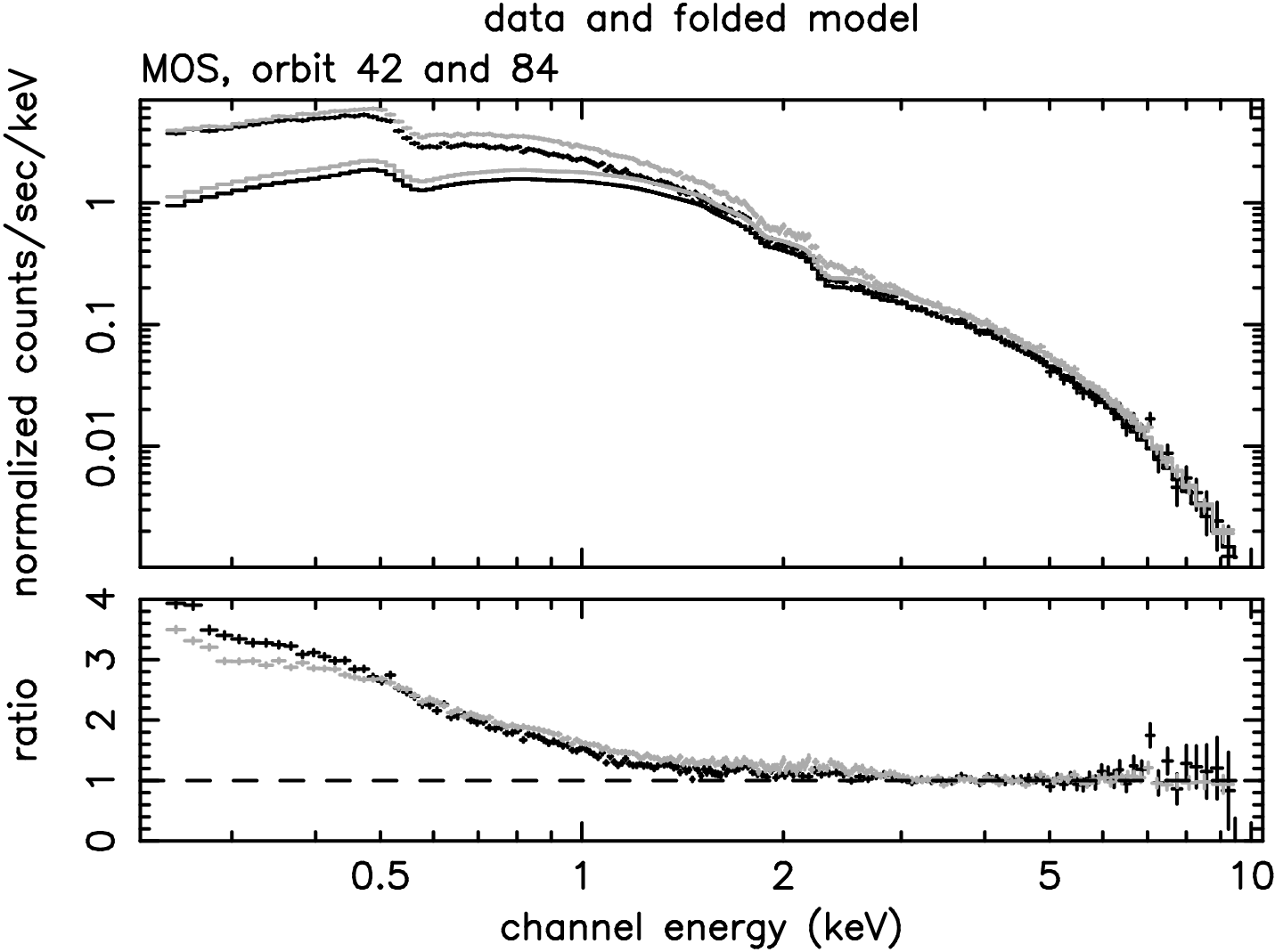}}}
\caption{The EPIC MOS spectra of PKS~0558$-$504 for
orbits 42 and 84. A power-law has been fitted to the 4--10~keV data
and extrapolated to lower energies. A broad, strong soft X-ray excess is
clearly seen extending to $\sim 3$ keV.}
\label{mosspec}
\end{figure}

As noted above, no large ($\ga 10$\%) changes in spectral shape were
observed during each orbit. Thus, to improve the signal-to-noise ratio
the MOS data were combined to form a mean spectrum for each orbit before
detailed spectral analysis.

The background-subtracted, summed MOS observations provide the highest
S/N data for PKS~0558$-$504 to date. These spectra are very poorly
fitted by a single power-law ($\chi^2_{\nu} = 2.75$) due to the
presence of a large, broad soft X-ray excess which 
dominates the X-ray emission from PKS~0558$-$504 in the {\it
XMM-Newton} bandpass. To illustrate this feature, Fig.\ \ref{mosspec}
shows the MOS spectra for each orbit with a power-law fitted to the
4--10 keV band, allowing for Galactic absorption (fit 1 in Table~1 --
all the fits are summarised in Table~1). The power-law is fitted above
4~keV as the soft excess appears to extend to $\sim 3$~keV. The PN
spectrum for orbit 84 is consistent with the MOS allowing for the
current calibration uncertainties. 

No evidence is found for a significant change in X-ray spectral shape
between the two orbits, but there is a small decrease in the mean X-ray
flux of $\approx 18$\% from orbit 42 to 84. The OM data
show that the UV flux also decreased by $\approx 10$\% from orbit 42 to
84. For orbit 42 the 0.2--2.0~keV and 2--10~keV fluxes are $2.44\times
10^{-11}$ and $1.01\times10^{-11}$ erg cm$^{-2}$ s$^{-1}$
respectively, corresponding to luminosities of $3.7\times10^{45}$ and
$0.98\times 10^{45}$ erg s$^{-1}$.

\begin{figure}
\resizebox{\hsize}{!}{\rotatebox{-90}{\includegraphics{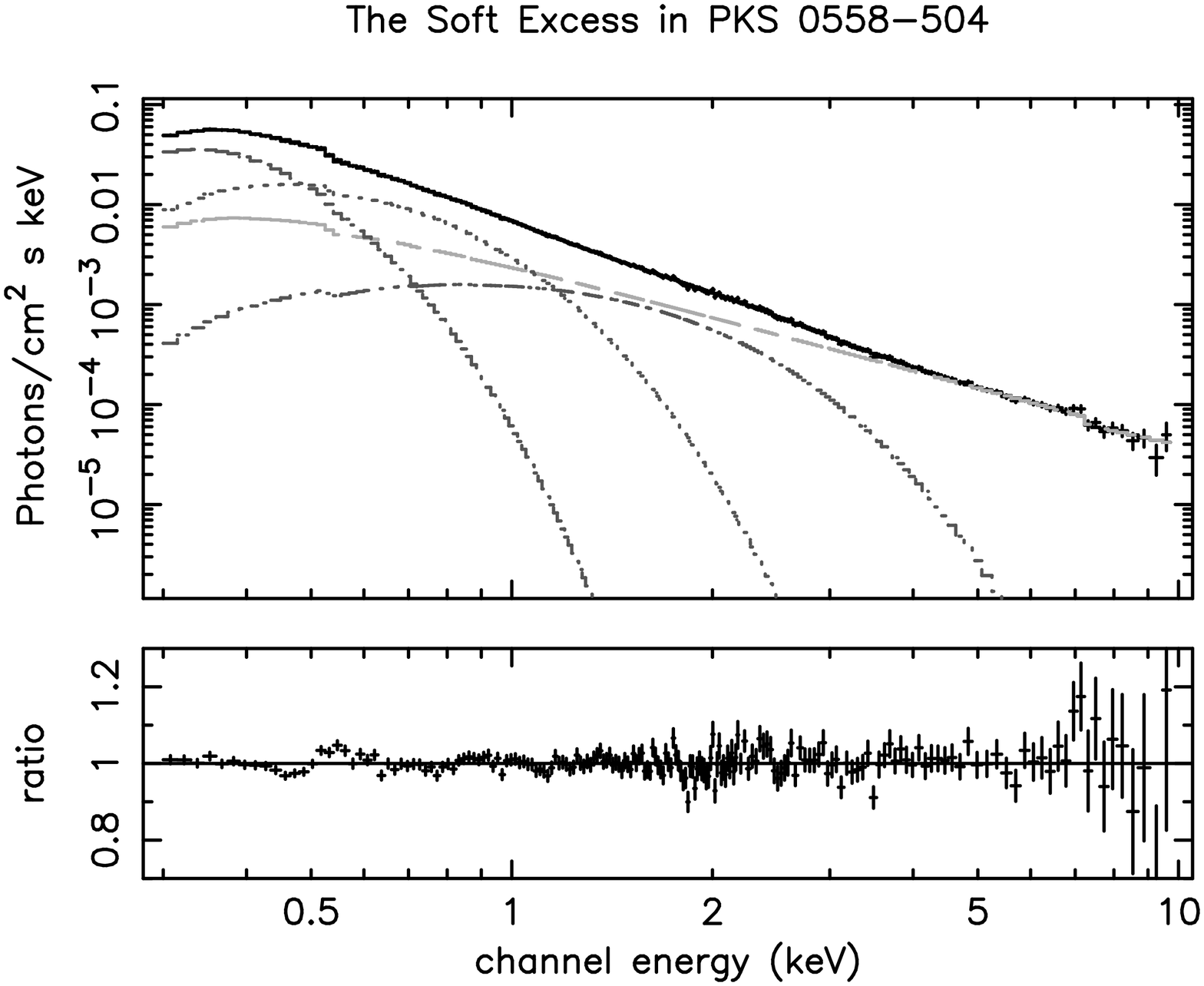}}}
\caption{The multiple blackbody plus power-law fit to the combined 
orbit 42 and 84 MOS data.}
\label{bbody}
\end{figure}

\begin{table*}
 \centering
 \begin{minipage}{140mm}
  \caption{Fits to {\it XMM-Newton} data in the 0.2--10 keV band. 
           $^a$Powerlaw index for fit to 4.0--10 keV. $^b$Electron
temperature of thermal source (eV). $^c$Electron temperature of Comptonizing
electrons (keV). $^d$Optical depth.}
  \begin{tabular}{@{}lllccr@{}}
   Rev/Camera & Fit & Model & $\Gamma$/photon kT & 
   BB or electron temp (kT)/optical depth & $\chi^2$/dof \\
\\
42/MOS & 1 & PL$^a$ & $2.12\pm0.15$ & -- & 241/243 \\
42/MOS & 2 & PL + 3$\times$BB & $2.03\pm0.09$& $82\pm5$, $173\pm10$,
$469\pm18$ & 608/489 \\
84/MOS & 1 & PL$^a$ & $1.99\pm0.15$ & -- & 276/258 \\
84/MOS & 2 & PL + 3$\times$BB & $1.97\pm0.12$& $74\pm5$, $165\pm11$,
$454\pm35$& 622/493 \\
84/PN  & 1 & PL$^a$ & $1.94\pm0.17$ & -- & 188/192 \\
84/PN  & 2 & PL + 3$\times$BB & $1.84\pm0.05$ & $89\pm5$, $189\pm14$,
$560\pm50$ & 1009/798 \\
42+84 MOS & 1 & PL$^a$ & $2.07\pm0.12$ & -- & 319/317 \\
42+84 MOS & 2 & PL + 3$\times$BB & $1.97\pm0.10$ & $80\pm3$, $175\pm6$,
$475\pm15$ & 721/556 \\
42+84 MOS & 3 & CompTT & $59\pm8$$^b$ & $4.0^{+4.5}_{-1.5}$$^c$, 
$2.5^{+0.8}_{-1.1}$$^d$ 
& 780/559 \\
& & & $59\pm8$$^b$ & $55^{+65}_{-35}$$^c$, $0.7\pm0.4$$^d$ & \\
\end{tabular}
\end{minipage}
\end{table*}

The soft X-ray excess in PKS~0558$-$504 is far too broad to be fitted
by a single temperature blackbody. To parameterise the spectra with a
relatively simple model, we fitted a combination of blackbodies and a
power-law. At least three blackbodies are required of 80, 175 and
475~eV, to give an acceptable fit ($\chi^2_{\nu} = 1.2$; fit 2 in
Table~1). Other than the relative normalisation (consistent with the
small overall change in flux), the derived power-law indices and
blackbody temperatures required to fit the mean spectra from orbits 42
and 84 are statistically consistent. Therefore, we combined the entire
MOS data before proceeding. The multiple blackbody and power-law fit
to the combined MOS data is shown in Fig.\ \ref{bbody}. The
high-energy power-law slope in the multi-component model is consistent
with the $\Gamma = 1.9$ commonly seen in radio-quiet AGN.

No iron fluorescence line is detected from either neutral or ionised
material. For the combined MOS data, the upper limits to the EW of a
6.4~keV line are $<10$~eV for a narrow line ($\sigma = 10$~eV) and
$<90$~eV for a broad line ($\sigma = 300$~eV). For a narrow ionised
line at 6.7~keV, the upper limit is 42~eV. In addition, no evidence is
found in the EPIC or RGS spectra for any intrinsic cold or warm
absorption. The {\ion{O}{vii}} or {\ion{O}{viii}} edge optical depth
limits are $\tau <0.1$. The upper limit for an ionised iron edge is
$\tau < 0.2$. We note that an apparent feature in Figs \ref{mosspec},
\ref{bbody} and \ref{compton} at $\approx 7$~keV (which is the wrong
energy for an iron line in the QSO rest-frame) is due to noise in some
of the MOS spectra that went to form the mean. This feature is not
present in the PN spectra. Overall, PKS~0558$-$504 is a powerful
quasar in the X-ray band, with a 0.25--10~keV (unabsorbed) luminosity
of $4.5 \times 10^{45}$ erg s$^{-1}$ mostly emitted below 3~keV. The
lack of spectral features strongly implies that we are seeing the bare
quasar continuum emission.

\section{Discussion}

\begin{figure}
\resizebox{\hsize}{!}{\rotatebox{-90}{\includegraphics{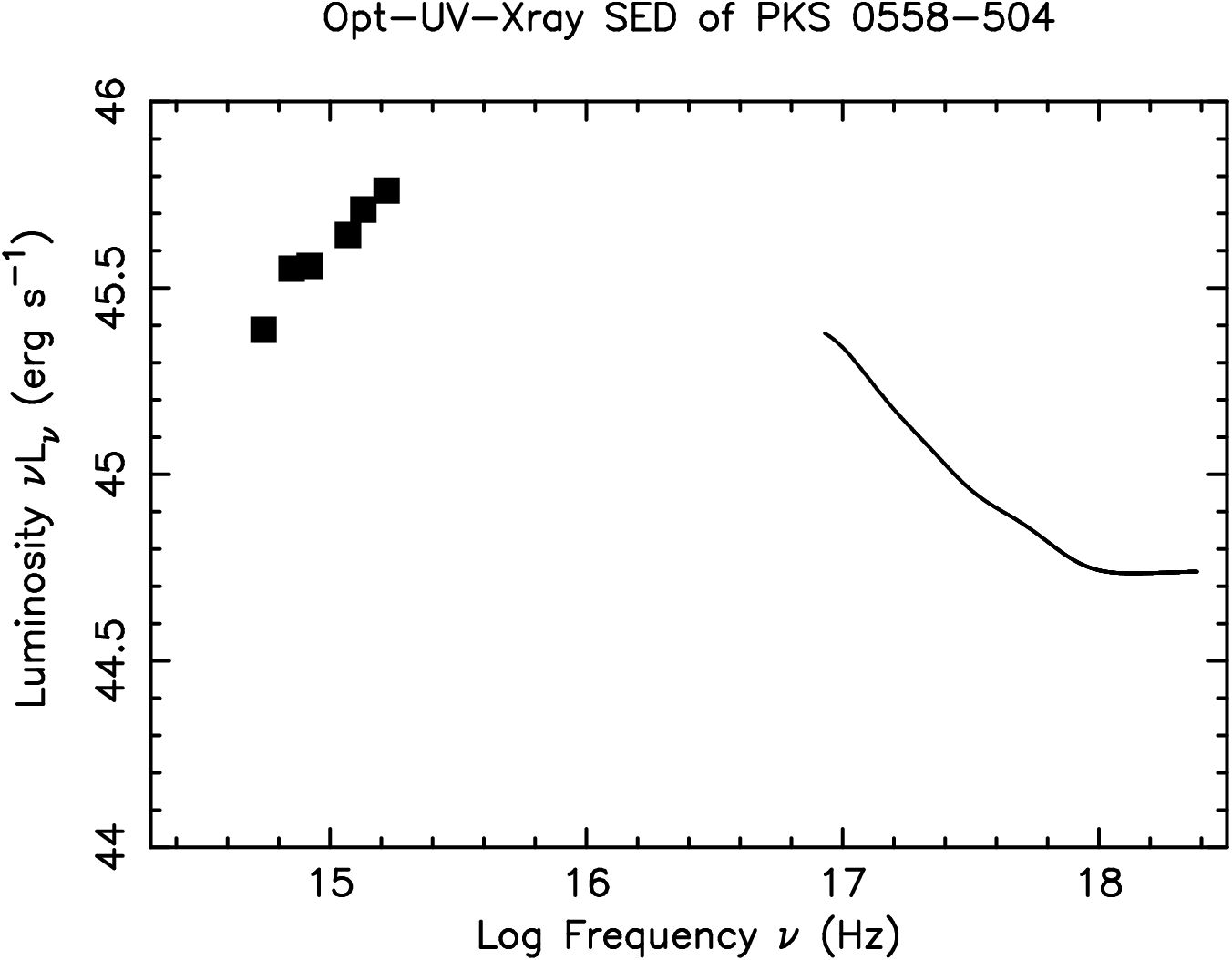}}}
\caption{The broad-band spectral energy distribution of PKS~0558$-$504
from OM and EPIC data.}
\label{sed}
\end{figure}

The multiple-blackbody plus power-law model discussed above provides an
acceptable fit to the EPIC spectra, but has no physical basis. To
examine the broad-band spectral energy distribution (SED) of
PKS~0558$-$504 the dereddened optical and UV data from the OM
(assuming Galactic E(B$-$V) $=0.044$) together with the X-ray
continuum derived from the combined MOS data (using the
multiple blackbody model) are shown in Fig.\
\ref{sed}.

The SED of PKS~0558$-$504 clearly shows an optical/UV BBB, and the
{\it XMM-Newton} data suggest this feature extends into the X-ray
band. To power an object as luminous as PKS~0558$-$504
($L_{\mathrm{bol}} \sim 10^{46}$ erg s$^{-1}$) would require a black
hole of mass $M_{\mathrm{BH}} \sim 10^8 $ $M_{\sun}$ accreting at the
Eddington rate. Based on accretion disc models (e.g. Czerny \& Elvis
\cite{cz}), the thermal spectrum of such a system could explain the
optical/UV portion of the BBB but would be unable to produce the
broad, soft X-ray excess or the high-energy power-law-like emission.

\begin{figure}
\resizebox{\hsize}{!}{\rotatebox{-90}{\includegraphics{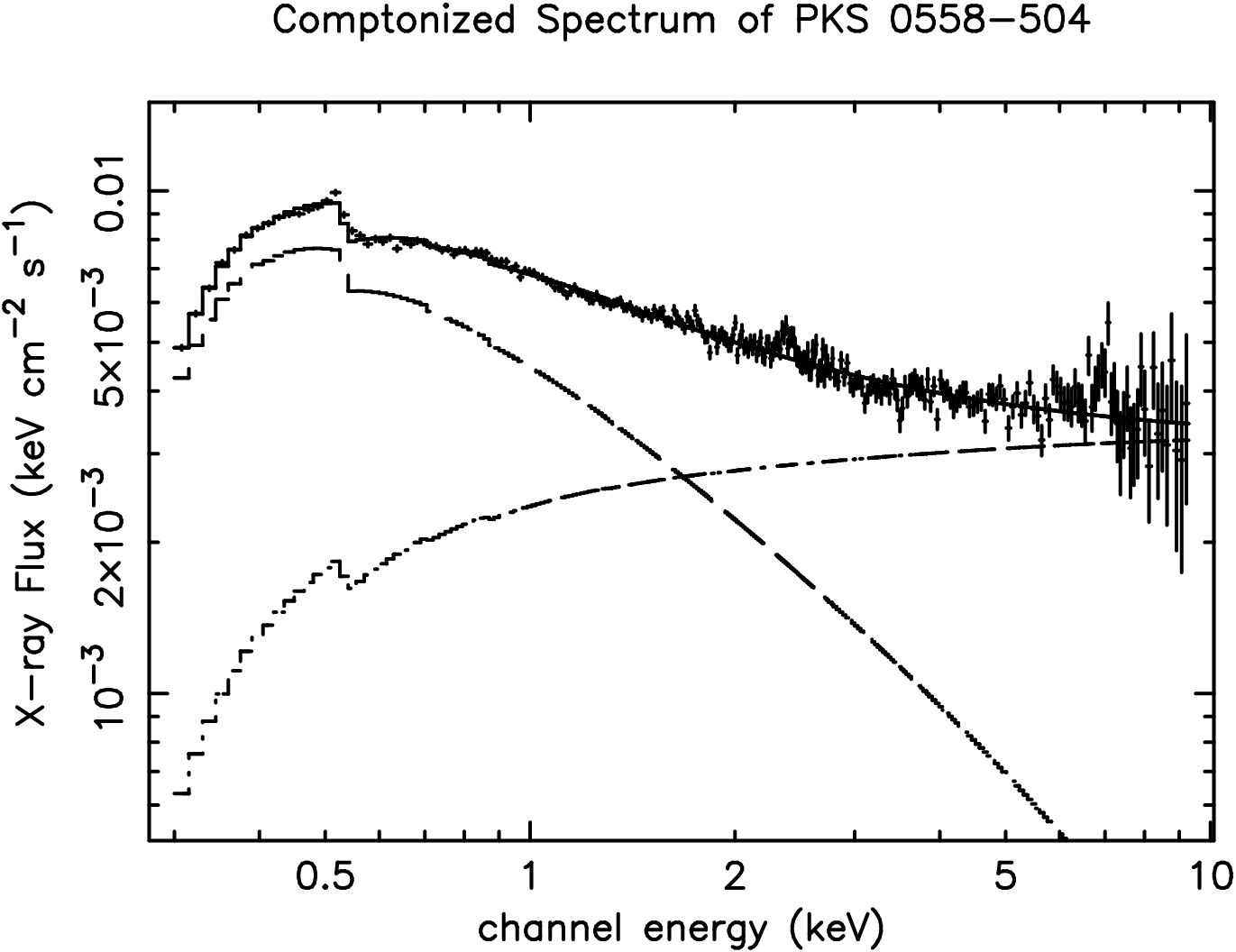}}}
\caption{The Comptonization model fit to the combined orbit 42 and
84 MOS data plotted in $\nu f_{\nu}$ space.}
\label{compton}
\end{figure}

One model suggested to explain the steep X-ray spectra of extreme soft X-ray
sources, such as NLS1s, is that these source accrete at an unusually
high rate leading to a large soft photon flux. These photons then Compton
cool the hard photon source, generated in a hot accretion disc corona,
resulting in a steeper X-ray spectrum (Pounds, Done \& Osborne \cite{po}).
The temperature, geometry and covering factor of disc coronae are
poorly defined from theory, but multiple-temperatures are likely to be
required due to vertical temperature stratification.

To explore Comptonization models, the results of fitting the {\sc
Xspec} model {\sc CompTT} to the combined orbit 42 and 84 data are
given in Table~1 (fit 3) and shown in Fig.\ \ref{compton}. A model
involving two electron populations, both scattering photons from a
single temperature thermal source, provides a good fit to the mean
X-ray spectrum. Allowing the thermal source temperature to vary
independently does not produce a significantly better fit. The
temperature and luminosity of the soft photon source are consistent
with that of a thermal accretion disc spectrum for a $10^8 M_{\sun}$
black hole accreting at close to the Eddington rate. The Comptonizing
electron temperatures of 4 and 55~keV are at the low end of the
expected temperature range (Haardt \& Maraschi \cite{ha}), possibly due to
the large soft photon flux Compton cooling the corona. We note that
the exact values of the parameters for the two electron populations in
the comptt model are quite poorly constrained. This is because
increasing the optical depth slightly can harden the spectrum in a
similar way to adding a higher temperature electron population, thus
giving a variety of two-component models that adequately fit the data.

We note that, as suggested by the power-law and multiple blackbody
fits, the high-energy tail of the Comptonized spectrum has a shape
consistent with that of `normal' broad-line AGN power-laws ($\Gamma
\sim 1.9$), and is significantly harder than that derived from the
{\it ASCA} data. These differences are presumably due to the inability
of the {\it ASCA} data to delineate the tail of the soft X-ray excess
which extends to around 3 keV in this source. The observed mean slope
of the hard power-law in AGN ($\Gamma = 1.9$) is predicted from models
in which feedback occurs between soft photons produced by hard X-ray
heating Compton cool a moderate optical-depth corona (Haardt \&
Maraschi \cite{ha}). Such a model cannot explain the bulk of the soft X-ray
luminosity in PKS~0558$-$504 as it dominates the luminosity, but may
produce the residual hard power-law emission.

The reduced chi-squared value for the Comptonized model ($\chi^2_{\nu} =
1.3$) shows that the Comptonization model fit leaves some spectral
curvature, mainly above a few keV. Some of this imperfection is due to
remaining calibration uncertainties around features such as detector Gold
edge, but most of it is due to the Comptonization model simply being too
simplistic to adequately fit such high-quality data. Some of the
remaining spectral curvature can be reduced by including an additional
Comptonizing component (i.e. hotter electrons) but this component does
not formally improve the fit in terms of the reduced chi-squared
presumably because the {\it XMM-Newton} high-energy cutoff is too low to
adequately constrain a very-hot electron population. 

An alternative explanation for a high-energy `power-law-like'
component is a small amount of reflection from an ionised disc. The
disc is likely to be ionised given the luminosity of the source.
Indeed, the absence of any strong iron K$\alpha$ emission line or iron
edge at high energies and the lack of absorption features due to
moderately ionised material at low energies suggests that any X-ray
heated disc will have a substantial `skin' which is fully ionised and
effectively acting as a `mirror'. This has been suggested for high
accretion rate sources in the context of a magnetic-flare model for
X-ray variability (Nayakshin \cite{na}). The small iron K$\alpha$ emission line
and iron edge strengths predicted by the magnetic-flare model, assuming
$\Gamma \sim 2$, are consistent with the upper limits from the {\it
XMM-Newton} spectra (Nayakshin, private communication). What is less
clear is if a fully ionised corona can be maintained in a source like
PKS~0558$-$504 with a large soft photon flux if those photons are
generated close to the magnetic-flares. Clearly the {\it XMM-Newton}
data for PKS~0558$-$504 demand comparison with more detailed accretion
disc models.

\section{Conclusions}

The {\it XMM-Newton} X-ray spectrum of PKS~0558$-$504 is dominated by
a large, variable soft excess. The fastest observed variability
implies a minimum efficiency of converting matter to energy of $\eta =
0.25$, suggesting accretion onto a Kerr black hole. Combining the
X-ray data with simultaneous OM optical and UV data reveal a large
big blue bump; too broad to be due purely to a thermal accretion disc
spectrum. The X-ray spectrum can be fitted by a Comptonization model
in which thermal photons, presumably from the disc, are scattered by a
hot corona. At least two electron temperatures are required. Overall,
PKS~0558$-$504 can be explained by an accretion disc running at a high
accretion rate producing copious amounts of soft EUV photons which are
then Compton cooling a hot disc corona.

\section*{Acknowledgements}

Based on observations obtained with the XMM-Newton, and ESA science
mission with instruments and contributions diectly funded by ESA
member states and the USA (NASA). EPIC was developed by the EPIC
Consortium led by the Principal Investigator, Dr. M. J. L. Turner. The
consortium comprises the following Institutes: University of
Leicester, University of Birmingham, (UK); CEA/Saclay, IAS Orsay, CESR
Toulouse, (France); IAAP Tuebingen, MPE Garching, (Germany); IFC
Milan, ITESRE Bologna, IAUP Palermo, (Italy). EPIC is funded by:
PPARC, CEA, CNES, DLR and ASI. We thank the EPIC and SSC teams for
much help and advice, particularly Steve Sembay, Gareth Griffiths,
Richard West and Ian Stewart. We also thank Kim Page for help with
data reduction.

\label{lastpage}

\end{document}